\documentclass[aps,twocolumn,groupedaddress,showkeys]{revtex4-1}
\usepackage{float}
\usepackage{multirow}
\usepackage{amsmath}
\usepackage{amssymb}
\usepackage{graphicx}
\usepackage{stackrel}
\usepackage{natbib}
\usepackage[unicode=true]{hyperref}

\begin{document}
	\title{Chandrasekhar limit for rotating quark stars}
	
	\author{Ashadul Halder}
	\email{ashadul.halder@gmail.com}
	\affiliation{Department of Physics, St. Xavier's College, 30, Mother Teresa Sarani, Kolkata-700016, India.}
	
	\author{Shibaji Banerjee}
	\email{shiva@sxccal.edu}
	\affiliation{Department of Physics, St. Xavier's College, 30, Mother Teresa Sarani, Kolkata-700016, India.}

	\author{Sanjay K. Ghosh}
	\email{sanjay@jcbose.ac.in}
	\affiliation{Department of Physics and Centre for Astroparticle Physics \& Space Science (CAPSS),\\
	Bose Institute, EN-80, Sector V, Bidhannagar, Kolkata-700091, India.}
	
	\author{Sibaji Raha}
	\email{sibaji.raha@jcbose.ac.in}
	\affiliation{Department of Physics and Centre for Astroparticle Physics \& Space Science (CAPSS),\\
	Bose Institute, EN-80, Sector V, Bidhannagar, Kolkata-700091, India.}
	
	\begin{abstract}		
		The limiting mass is a significant characteristic for compact exotic stars. In the case of quark stars, the limiting mass can be expressed in terms of fundamental constants and the Bag constant. In the present work, using bag model description, the maximum mass of a rotating quark star is found to depend on the rotational frequency apart from other fundamental parameters. The analytical results obtained agree with the results of several relevant numerical estimates as well as observational evidences.
	\end{abstract}
	\keywords{Quark Star; Compact Star; Limiting mass}
	\pacs{}
	\maketitle

	\section{Introduction}
		The theoretical investigation for quark stars has turned out to be a worldwide enterprise after the first prediction of the quark core at the center of the core collapsed neutron stars \cite{ivan_1}. The idea of the quark star was brought to light by soviet physicists Ivanenko and Kurdgelaidze \cite{ivan_1,ivan_2}, about five years after the Gell-Mann prediction of quarks \cite{Quark_Wikipedia}. Several recent observations of compact and comparatively cooler stars (SWIFTJ1749.4-2807 \cite{hyu}, RXJ185635-3754 and 3C58 \cite{Prakash:2002xx}) also provide evidence in favor of the existence of quark stars. Unlike other compact astrophysical objects and main sequence stars, the quark stars are self-bound by strong interaction rather than by gravity alone  \cite{Jaikumar:2004zy,Witten}. Apart from this unique characteristic, the density of the quark stars is remarkably high (even higher than nuclear density i.e. $2.4\times10^{14}\,\rm{g/cm^{3}}$ \cite{Bag_Model_2}) and hence gain immense importance as natural laboratory for quark matter.
		
		The quark star is considered to be the very end product of the stellar evolution. As the fuel of a star (neutron star) tends to get exhausted, the radiation pressure of the star can no longer balance the self-gravity. As a result, the core part of the star starts collapsing due to its own gravity. However, in certain cases, the density turns out to be substantially high and therefore a quark core is spontaneously generated (even without strangeness) at the center of the star. Although initially, the collapse takes place in the core part, it grows over time and occupies the entire star eventually by capturing free neutrons from the vicinity of the surface in absence of the Coulomb barrier. Besides this process, a thin outer crust of comparatively lower density is also generated outside the quark core ($\sim 1/1000$ of the average density of the star), which does not contain any free neutrons \cite{Witten}. Consequently, the effect due to the outer crust of the quark star is ignorable in limiting mass as well as the limiting radius. The newly formed quark phase contains only 2 flavors of quarks (`u' and `d' quark) as it is generated from the non-hyperonic baryons (mainly neutron). Later the strangeness can be developed in the 2 flavor quark state via weak interaction ($u d \rightarrow u s$, in excess of `d' quark) by absorbing the energy of the quark star \cite{ns_to_qs, alcock} and thus transforms into a strange quark star (SQS). The SQS is basically composed of 3 flavors of quarks (contains strange quark in addition to the `u' and `d' quark) confined in a hypothetical large bag \cite{Bag_Model_1,Bag_Model_2}, which is characterized by the Bag Constant. The strange quark matter can be treated as the perfect ground state of the strongly interacting matter as predicted by E. Witten \cite{Witten,Jaikumar:2004zy}. Also, there are several alternative models \cite{Strange_qs_1,Strange_qs_2,weber_structure} supporting the conjecture of the SQS (for a review, see \cite{Strange_Q_Star}). But in another possibility, a quark star can be formed by the clumping of ambient quark matters due to the gravitational interaction essentially like other normal stars. This is only possible if a significant percentage of strange quarks \cite{Witten} present in the system. Such ``pure'' quark stars \cite{Pure_qs} might belong to a hypothetical quark galaxy or might result from the accumulation of strangelets \cite{Quark_galaxy,qg1,qg2}. Primordial strange star can be another possible example of the quark star. According to this conjecture, quark stars were formed due to the phase transition in the early Universe. Later those transform into strange stars in order to maintain stability as assumed in the work of \cite{Witten}.
		
		The key difference between the ordinary stars and the quark star is that the mass of an ordinary star is almost entirely due to the baryons, whereas for the quark star, one has to define the effective mass for the `u' and `d' quarks, as those (quarks) are known to have very small masses. As the quark stars may be expected to be produced due to the hadron quark transition in neutron stars, its limiting mass is expected to be very close to that of the neutron star. As a result, one can estimate the total mass of the star by solving the Tolman-Oppenheimer-Volkov (TOV) equations \cite{tov} numerically \cite{lm1,lm2} by adopting the conjecture, proposed by E. Witten \cite{Witten}. In several recent works \cite{rotlm,rotlm_2}, the limiting mass for spinning quark star is also studied using the same procedure as used for the case of static quark star. But there is no argument in the literature that favors the existence of limiting mass which, like ordinary compact objects \cite{shapiro}, depends dominantly on the fundamental constants. 
		Moreover, unlike neutron stars (which are bound by gravity), quark stars are self-bound by strong interaction (see Ref.~\cite{Jaikumar:2004zy,def_agnst}). Thus, it is not strictly necessary to include the density as a function of radius. Quark stars are confined compact objects, but their densities hover near nuclear densities and hence the bag constant seems to cover appropriately the strong interaction section, particularly in view of the density dependent modeling (QMD) of the quark star masses. A salient feature of this approach is that it is independent of any particular EOS (equation of state) models and the entire dynamics comes from the confinement mechanism modeled by QMD.
		In the work of Banerjee {\it et. al.} \cite{SBa}, an analytical study for the limiting mass carried out for static quark stars, starting from the energy balance relations as proposed by Landau \cite{Landau}, is found to depend mainly on the fundamental constants and the MIT bag constant \cite{Bag_Density}. In the current work, we look into such a theoretical limit for the case of rotating quark stars and estimate the possible limiting frequency of such compact stars. The effect of the bag constant and rotational frequency at the limiting mass and radius has also been studied in the theoretical regime.
		
		Since, the quark stars are self-bound by strong interaction \cite{Jaikumar:2004zy,def_agnst} (which is far stronger than the gravitational field), it can withstand high rotational frequency without having notable shape deformation (ellipticity $\lesssim 10^{-4}-10^{-7}$ \cite{def_sphere_1,ushomirsky2000} even in the case for millisecond (time period) stars). So, the modeled quark stars are considered to be rigid and spherical in our entire calculation. But as those stars are highly dense and massive, the volume of such a compact star can no longer remain an Euclidean sphere (i.e. volume $\frac{4}{3}\pi \rm{Radius}^3$) due to the curvature of space-time. So, the general relativistic correction is to be taken into account in the calculation of the volume of the star. 		
		
		The paper is organized as follows, in the sections~\ref{sec_bag} and \ref{sec_ef}, we give an account of the bag constant and  the fermi energy respectively. In section~\ref{sec_emass}, the effective mass per particle of the quark star is calculated in the context of particle physics. Section~\ref{sec_energy} deals with the energy calculation and the results. Finally in section~\ref{sec_conc}, concluding remarks are given.
		
	\section{Bag Constant \label{sec_bag}}
		MIT Bag model has turned out to be a successful model of the hadronic structure and achieved immense successes in hadron spectroscopy \cite{bag_sp1,bag_sp2,bag_sp3}. 
		According to this phenomenological model, massless point-like quarks are confined in a hypothetical bag and the state is characterized by a parameter bag constant $\mathcal{B}$.

		The bag constant $\mathcal{B}$ depicts the difference between the vacuum energy densities of the non-perturbative and the perturbative ground states of the quarks \cite{Bag_Density} and depends on density (and temperature, in general).
		In the case of quark star, the entire star is assumed to be such a bag, which contains all the constituent quarks.
		
		The primary objective of this paper is to obtain analytical results of a rotating star that can be compared relative to a static star. In order to achieve this, we have made use of only the essential features of this model, to the extent of providing a dynamic mass to the constituent light quarks. This approach has also been adopted in several other recent works \cite{Strange_qs_1,mmhs,Bag_Model_2} which address the issue of the static mass limit of similar compact stars.

	\section{Fermi Energy \label{sec_ef}}
		In the current work, the limiting mass for the quark star is estimated by adopting the simple energy balance picture, as proposed by Landau \cite{Landau}. According to this approach, the total energy per fermion ($e$) attains the minima at the limiting case of the star, given by,
		\begin{equation}
			e=e_f+e_G+e_{rot},
		\end{equation}
		where, $e_f$, $e_G$ and $e_{rot}$ represent the contributions of the fermi energy, gravitational energy and the kinetic energy due to the rotation of the star respectively.  
		The Fermi energy density of the non-interacting fermions measures the maximum occupied energy by the fermions per unit volume, at the ground state of the system \cite{Fermi_Hyperphysics}. In the case of the quark star, the fermi energy occupies a significant fraction of the total energy of the star, mainly depending on the fermi number density ($n$) and radius of the star ($R$). The fermi energy density can be expressed in terms of chemical potential ($\mu$), given by \cite{SBa}, 
		\begin{equation}
			\mathcal{E}_{f}=\frac{g}{8\pi^{2}}\mu^{4},\label{eq:1}
		\end{equation}
		where, $g$ is the statistical degeneracy factor of the system. The chemical potential ($\mu$) for a star having $N$ number of fermions is described as,
		\begin{equation}
			\mu=\left(\frac{9\pi}{2g}\right)^{\frac{1}{3}}\frac{N^{\frac{1}{3}}}{R}.\label{eq:2}
		\end{equation}
		So, the expression for the Fermi energy per particle takes the form,
%
		\begin{equation}
			e_{f}=\frac{\mathcal{E}_{f}}{n}=\frac{3}{4}\left(\frac{9\pi}{2g}\right)^{\frac{1}{3}}\frac{N^{\frac{1}{3}}}{R}\label{eq:3}.
		\end{equation}
	
	\section{Effective mass per particle \label{sec_emass}}
		The effective mass of the quarks is an essential quantity, in order to estimate the total mass of the star as well as the gravitational potential and the rotational kinetic energy. According to our assumption (rigid and spherical star), the effective mass of the entire star ($M$) is given by,
		\begin{equation}
			M=e_{f}N+V\mathcal{B}.\label{eq:MM}
		\end{equation} 
		where $V$ is the volume of the star.
		Extremizing the above expression of total mass $M$ with respect to $R$, and further simplifying, the expression for bag constant ($\mathcal{B}$) is reduced.
		\begin{equation}
			\mathcal{B}=\frac{e_f N}{3 V},\label{bag}
		\end{equation}
		Applying the above expression of bag constant ($\mathcal{B}$) in Equ.~\ref{eq:MM}, the simplified form of the total mass ($M$) of the star is obtained, given by
		\begin{equation}
			M=4V\mathcal{B},\label{eq:total mass}
		\end{equation}
 		and thus the effective mass ($m$) of each quark inside the star,
		\begin{equation}
			m=\frac{M}{N}=\frac{4V\mathcal{B}}{N}.\label{eq:12}
		\end{equation}

		Being fermions, the effective mass of the quark coincides with the quark chemical potential. As a result, from the limit of vanishing quark density \cite{Fowler_Raha_Weiner,Fowler_Raha_Weiner_1,SBa} we get, 
		\begin{equation}
			\begin{array}{cccc}
				&\mu&=&\mathcal{B}/n\\
				or,&n&=&\mathcal{B}/\mu.
			\end{array}\label{eq:13}
		\end{equation}
		Now applying Eq.~(\ref{eq:2}) and (\ref{eq:13}) in Eq.~(\ref{eq:12}), the desired expression for effective mass per quark particle ($m$) can be obtained in terms of fundamental constants and bag constants.

	\section{Formalism and Result \label{sec_energy}}
		The gravitational potential and rotational kinetic energy per particle at the point $(R,\theta,\phi)$ (in spherical polar coordinate system) with respect to the center of the star, can be written in terms of effective mass and other parameters as,
		\begin{equation}
		\begin{array}{ll}
			e_{G}=-G\dfrac{Mm}{R} &\rm{   and}\\
			e_{\rm{rot}}=mc^2(\gamma-1).
		\end{array}
		\end{equation}
		 where $\gamma=\dfrac{1}{\sqrt{1-(\frac{2\pi\omega R\:cos\:\theta}{c})^{2}}}$ and $c$ is the speed of light in space. From the above expression it can be observed that, unlike the gravitational and fermi energy, the kinetic energy per particle changes over $\theta$ for any fixed value of $R$. So, in order to overcome the $\theta$-dependence in further calculations, a $\theta$-averaged  kinetic energy term ($\langle e_{\rm{rot}} \rangle$) is taken into account. Now the total effective energy per particle takes the form
		\begin{equation}
			e=e_{f}+e_{G}+\langle e_{\rm{rot}} \rangle.\label{eq:16}
		\end{equation}
		According to the Landau's energy balance picture, the limiting radius ($R_{\text{max}}$) and the corresponding mass ($M_{\text{max}}$) of the star can be obtained by extremizing the total energy per particle ($e$) with respect to the total number of particles ($N$). Although the star is assume to be spherical in our entire calculation, the volume $V$ of the star is not same as that of an Euclidean sphere (i.e. $\frac{4}{3}\pi R^3$) as those stars are highly dense and massive \cite{synge}. As a result, the general relativistic correction in volume is not negligible in the present calculation, hence we use relativistic volume \cite{hobson}, given by
		\begin{equation}
			dV = \sqrt{|g_{\mu \nu}|}  dr d\theta d\phi,
		\end{equation}
		where, $g_{\mu \nu}$ represents the $\mu \nu^{\rm{th}}$ component of the space-time metric and ($r,~\theta,~\phi$) are the coordinate in spherical polar coordinate system.

\begin{figure*}
			\centering{}
			\begin{tabular}{cc}
				\includegraphics[trim=15 30 20 50, clip,width=\columnwidth]{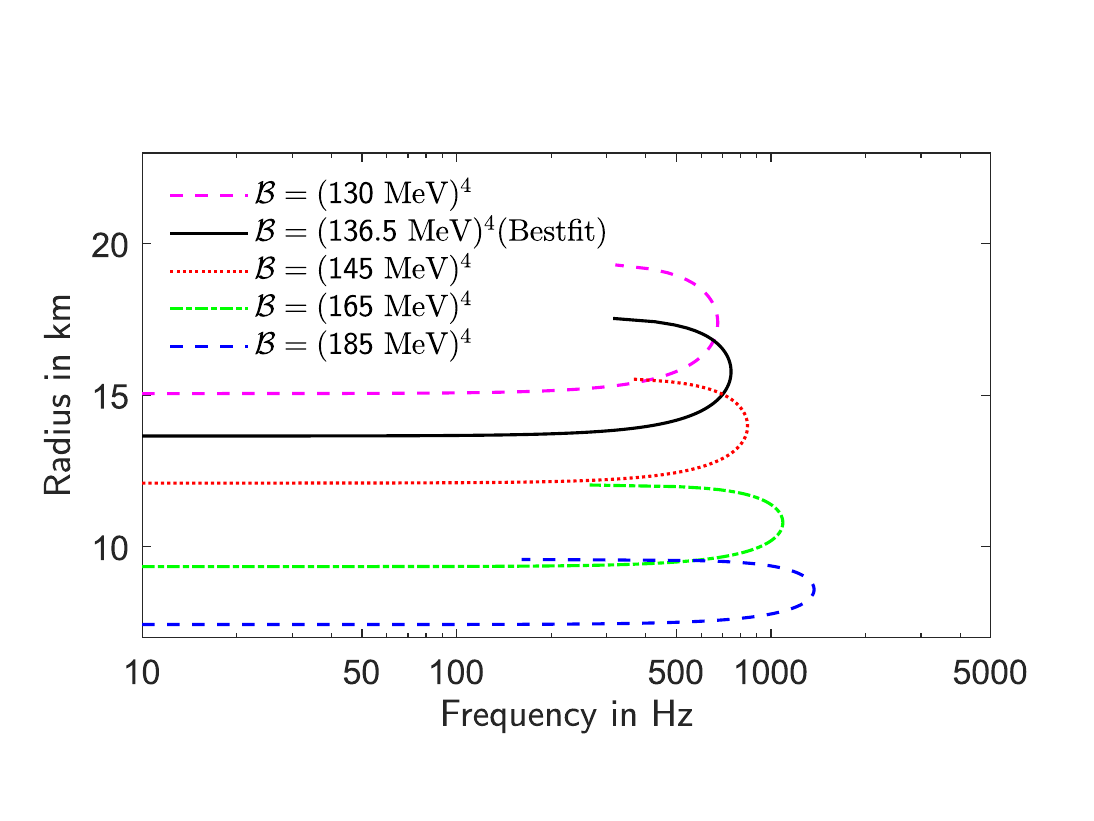} &
				\includegraphics[trim=20 30 20 50, clip,width=\columnwidth]{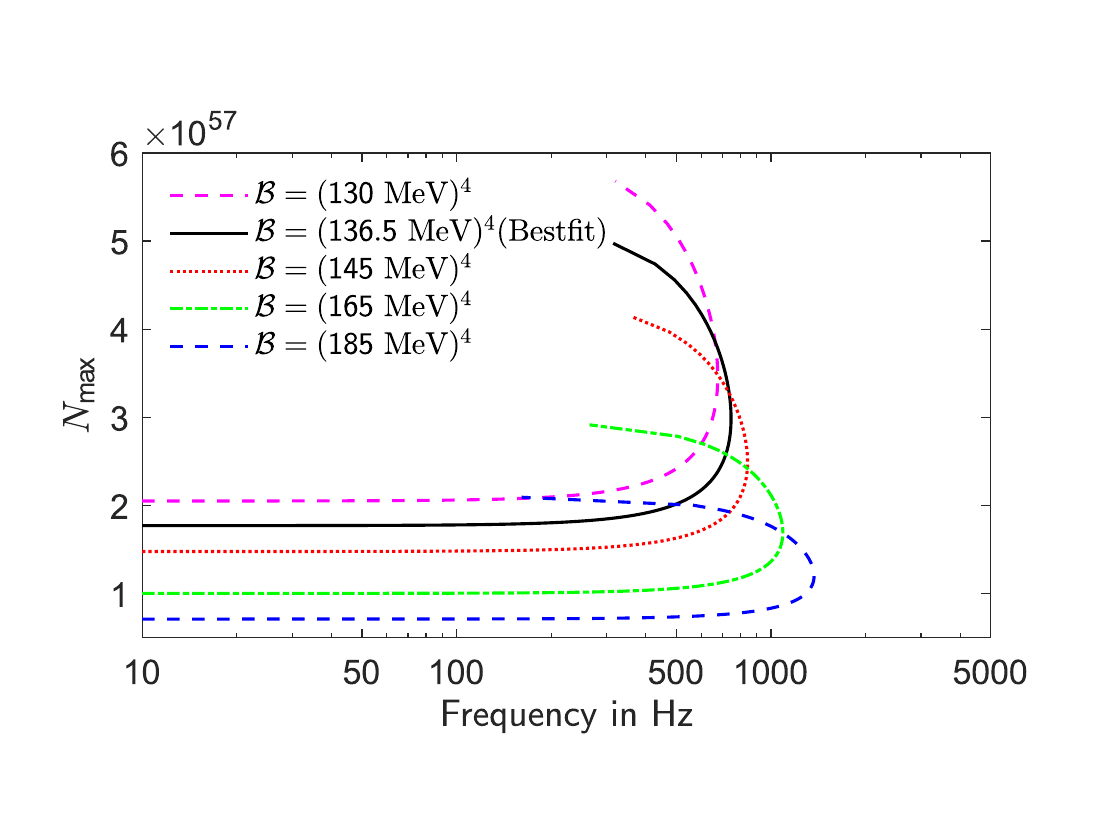}\\
				(a) & (b)\\
			\end{tabular}
			\begin{tabular}{c}
				\includegraphics[trim=0 30 20 50, clip,width=\columnwidth]{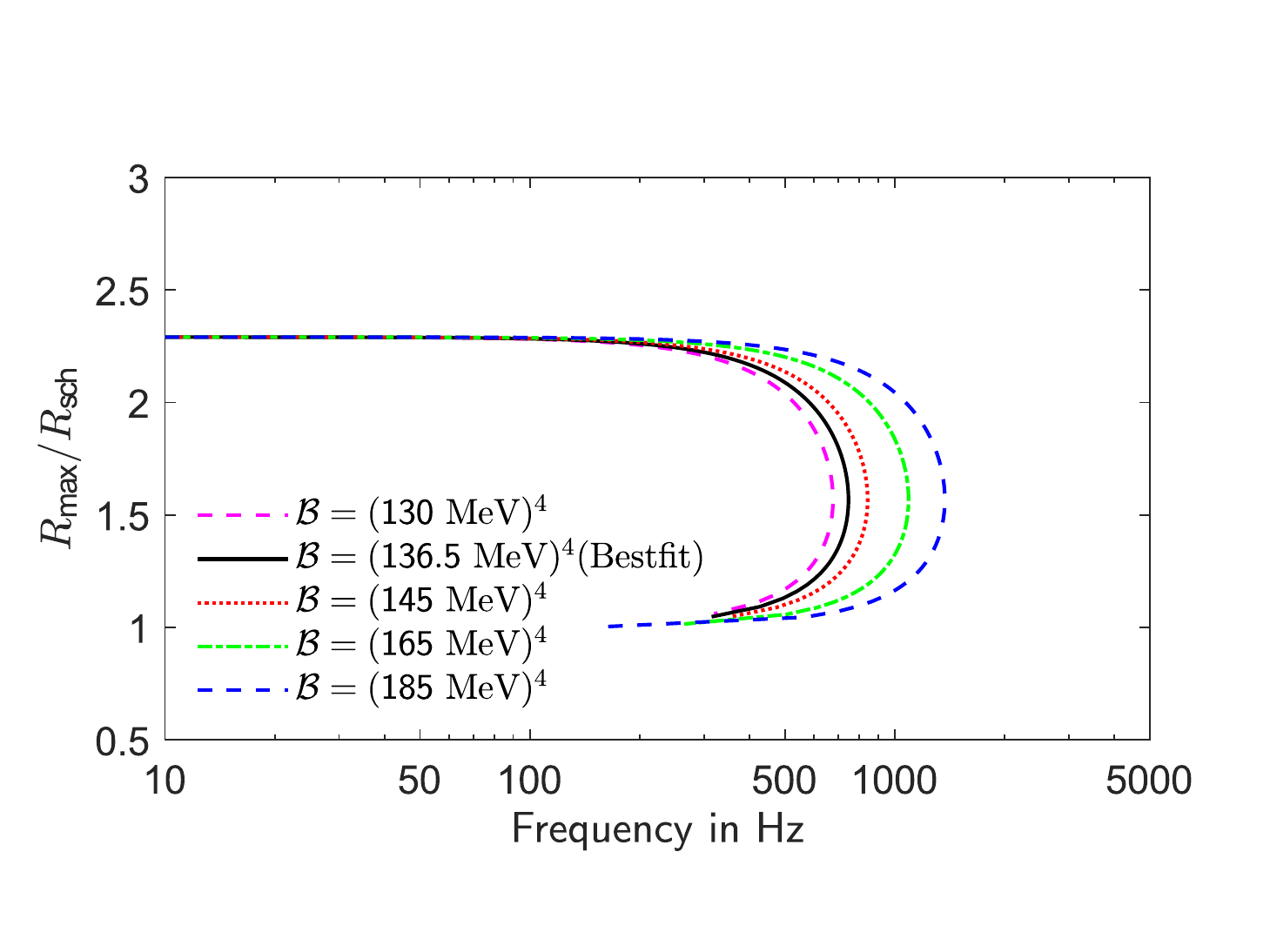}\\
				{(c)}\\
			\end{tabular}
			\caption{\label{fig:max nu}Variation of (a) limiting radius ($R_{\text{max}}$), (b) $N_{\text{max}}$ and (c) Radius mass ratio with $\nu$ for different values of bag constant.}
		\end{figure*}

		The limiting mass and radius, which are calculated by extremizing the total energy $e$, is independent of the degeneracy factor $g$ (see Ref.~\cite{SBa}, the result holds also in the rotating case). Consequently, we address a general solution for both types of quark stars (2-flavored normal quark star and 3-flavored strange quark star). In the current work, we also studied the dependence of the limiting mass, radius and the total number of particles $N$ on the bag constant as well as the rotational frequency.
		\begin{figure}
			\centering{}
			\includegraphics[trim=20 30 20 55, clip,width=\columnwidth]{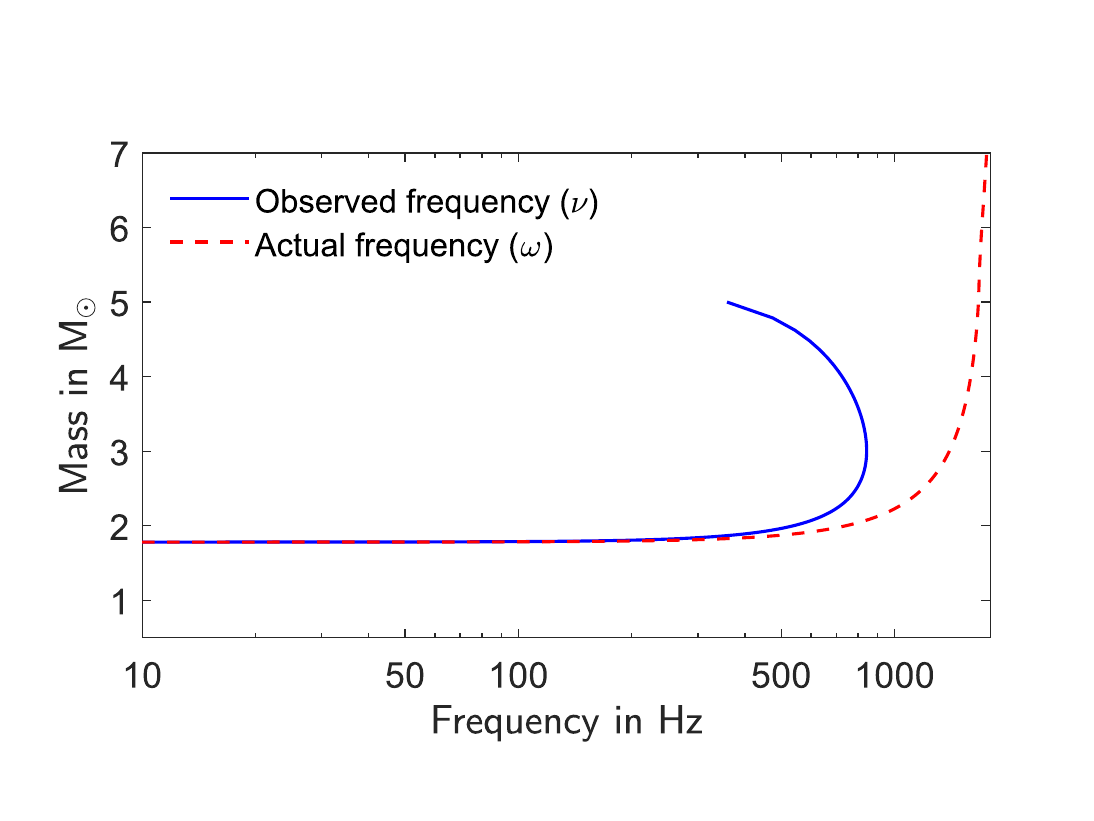}
			\caption{\label{fig:mmax vs freq2} Variation of the limiting mass ($M_{\text{max}}$) with the rotational frequency the star. The solid line represents the variation for the actual frequency $\omega$ while the dotted line is for the observed redshifted frequency ($\nu$)}
		\end{figure}
		The rotational effect of the star is parameterized by the rotational frequency ($\omega$), however, due to the extreme compactness and mass of the star (radius to mass ratio $\leq 2.2903$ in geometric unit, see Fig.~\ref{fig:max nu}c), the general relativistic effects emerge prominently in the observed rotational frequency ($\nu$) (as observed by a far away observer) \cite{gr_freq_2,gr_freq_1}. As a result, a far away observer would observe a red shifted form ($\nu$) of the actual frequency ($\omega$), given by
		\begin{equation}
			\nu=\omega \sqrt{1-\frac{2 G M}{c^2 R}},
		\end{equation}
		where, $M$ and $R$ are the mass and radius of the star respectively.
		Consequently, a significant deviation from $\omega$ is obtained in the observed frequency ($\nu$) in the case of fast spinning stars as shown in (see Fig.~\ref{fig:mmax vs freq2}). The variation of the limiting mass $M_{\rm{max}}$ with observed red-shifted frequency ($\nu$) and actual frequency ($\omega$) is described in Fig.~\ref{fig:mmax vs freq2} for a chosen value of bag constant $\mathcal{B}=(145~\rm{MeV})^4$.
		
		\begin{table}
			\centering{}
			\begin{tabular}{l c l}
				\hline \hline
				Object & Mass in $\rm{M}_{\odot}$ & Ref.\\
				\hline
				PRS J1614-2230 & $1.928^{0.017}_{0.017}$ & \cite{J1614-2230, Demorest}\\
				PSR J0348+0432 & $2.01^{0.04}_{0.04}$ & \cite{J0348+0432}\\
				PSR J0740+6620 & $2.14^{+0.1}_{-0.09}$ & \cite{J0740+6620}\\
				PSR J1311-3430 & $2.15-2.7$ & \cite{J1311-3430}\\
				PSR B1957+20 & $2.4^{0.12}_{0.12}$ & \cite{B1957+20}\\
				PSR J1600-3053	& $2.3^{+0.7}_{-0.6}$ & \cite{J1600-3053_1,J1600-3053}\\
				PSR J2215+5135	& $2.27^{+0.17}_{-0.15}$ &	\cite{J2215+5135_1,J2215+5135}\\
				PSR J0751+1807 & $2.10^{0.2}_{0.2}$ & \cite{J0751+1807,J0751+1807_1}\\
				PSR B1516+02B & $1.94^{+0.17}_{-0.19}$ & \cite{B1516+02B}\\
				\hline \hline
			\end{tabular}
			\caption{\label{tab:stars} Newly discovered massive pulsars.}
		\end{table}
		
		\begin{figure*}
			\centering{}
			\includegraphics[trim=0 40 30 60, clip,width=0.9\textwidth]{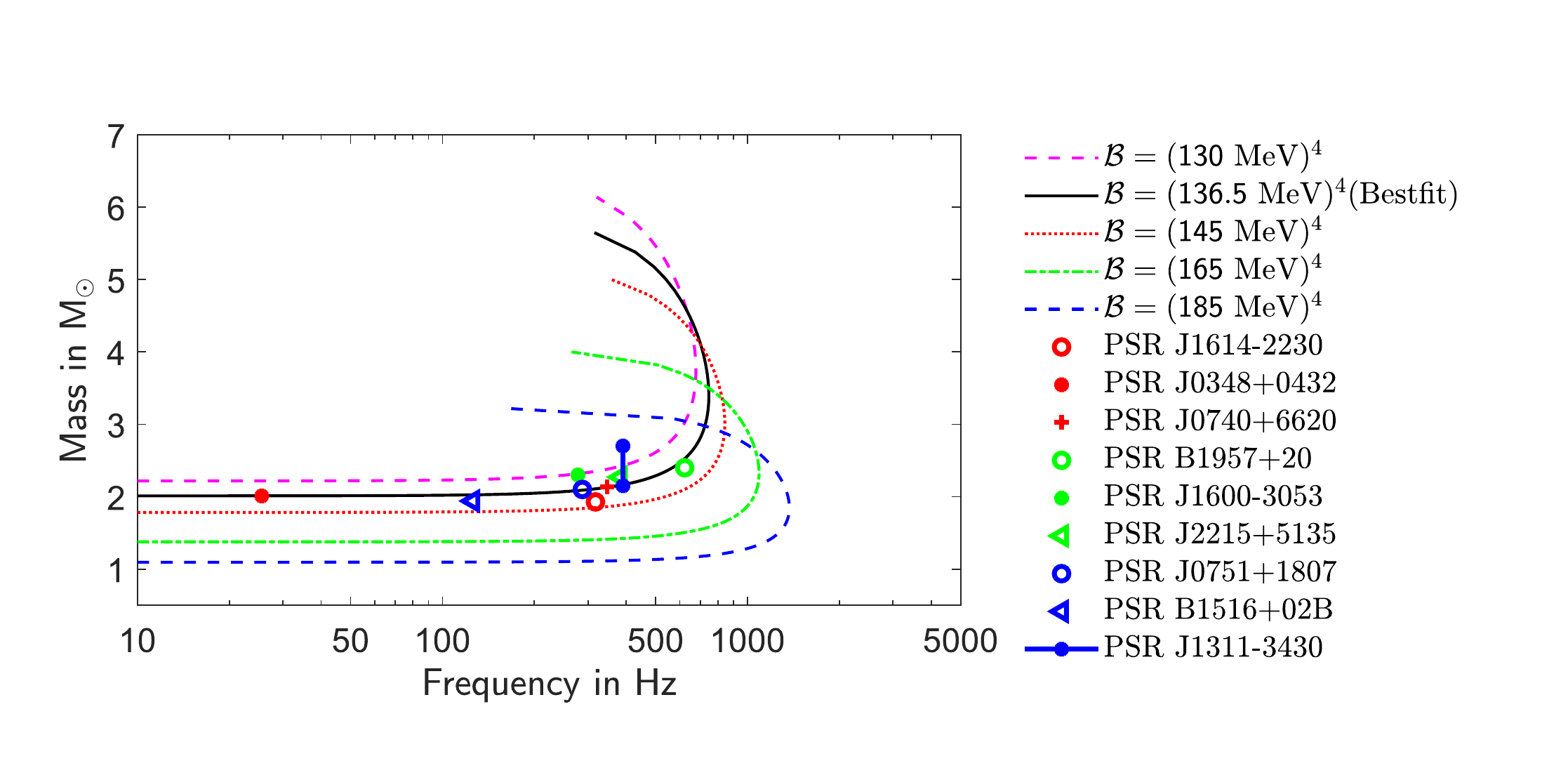}
			\caption{\label{fig:mmax nu} Variation of limiting mass ($M_{\text{max}}$) with observed frequency ($\nu$) for different values of bag constant ($\mathcal{B}$).}
		\end{figure*}
		
\begin{figure}
			\centering{}
			\includegraphics[trim=0 40 30 60, clip,width=\columnwidth]{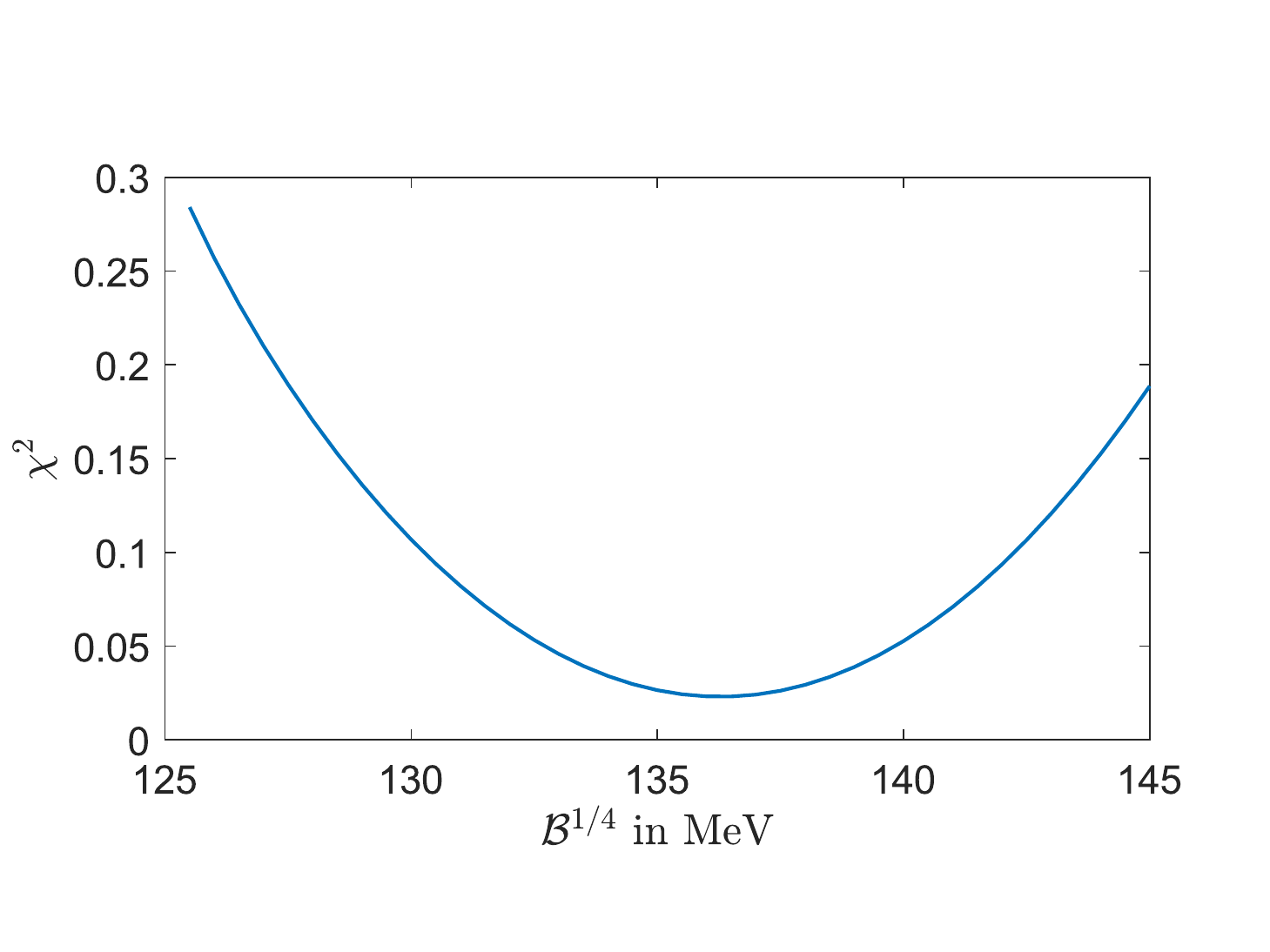}
			\caption{\label{fig:chi2} Variation of $\chi^2$ value with bag constant $\mathcal{B}$.}
		\end{figure}

		From Fig.~\ref{fig:mmax nu} it is evident that, for each chosen value of bag constant, the limiting mass remains almost unchanged in the lower frequency range. But as the frequency ($\nu$) reaches $\sim 300$ Hz, the kinetic energy becomes sufficiently high to be taken into consideration for the evaluation of the effective mass. As a result $M_{\text{max}}$ starts increasing gradually with frequency, however it suffers a rapid increase above the frequency of 600 Hz. Beyond a certain limit of $M_{\text{max}}$ for a given bag constant ($\mathcal{B}$) the star becomes so massive that observed frequency ($\nu$) starts falling with increasing $\omega$ due to gravitational effect. This provides a limiting value of $\nu$ (i.e. $\nu_{\text{max}}$) (see Fig.~\ref{fig:mmax nu}) for each value of bag constant. We also show the masses and observed frequencies of some recently discovered massive compact stars (see Table~\ref{tab:stars}) in the same plot and compared with our calculated masses and frequencies for different values of bag constants (using $\chi^2$ analysis). It is to be noted that, the stars (pulsars) mentioned in Table~\ref{tab:stars} are probably neutron star with or without quark core, rather than a pure quark star. But since their average densities are estimated to be close to the nuclear density, one can compare those pulsars with the calculated outcomes of the current work. The $\chi^2$ for our fit is defined as $\chi^2=\sum_{i}\frac{M_{\rm{cal}}-M_{\rm{obs}}}{M_{\rm{cal}}}$, where, $M_{cal}$ represents the calculated mass and $M_{obs}$ is the mass estimated from the observation (Table~\ref{tab:stars}). From our $\chi^2$ analysis the best-fit value of $\mathcal{B}$ is obtained around (136.5 MeV)$^4$ (see Fig.~\ref{fig:chi2}. PSR J1311-3430 is excluded for larger uncertainty in mass). For this best fitted value of bag constant, the mass variation with observed frequency is shown by the solid black line in Fig.~\ref{fig:mmax nu}. Fig.~\ref{fig:max nu}a and Fig.~\ref{fig:max nu}b describe the variation of limiting radius $R_{\text{max}}$ and corresponding number of particle containing the star ($M_{\rm{max}}$) respectively. In both the case, the same limiting frequencies ($\nu_{\text{max}}$) are observed for individual values of bag constants.

\begin{figure}
			\centering{}
			\includegraphics[trim=0 40 30 60, clip,width=\columnwidth]{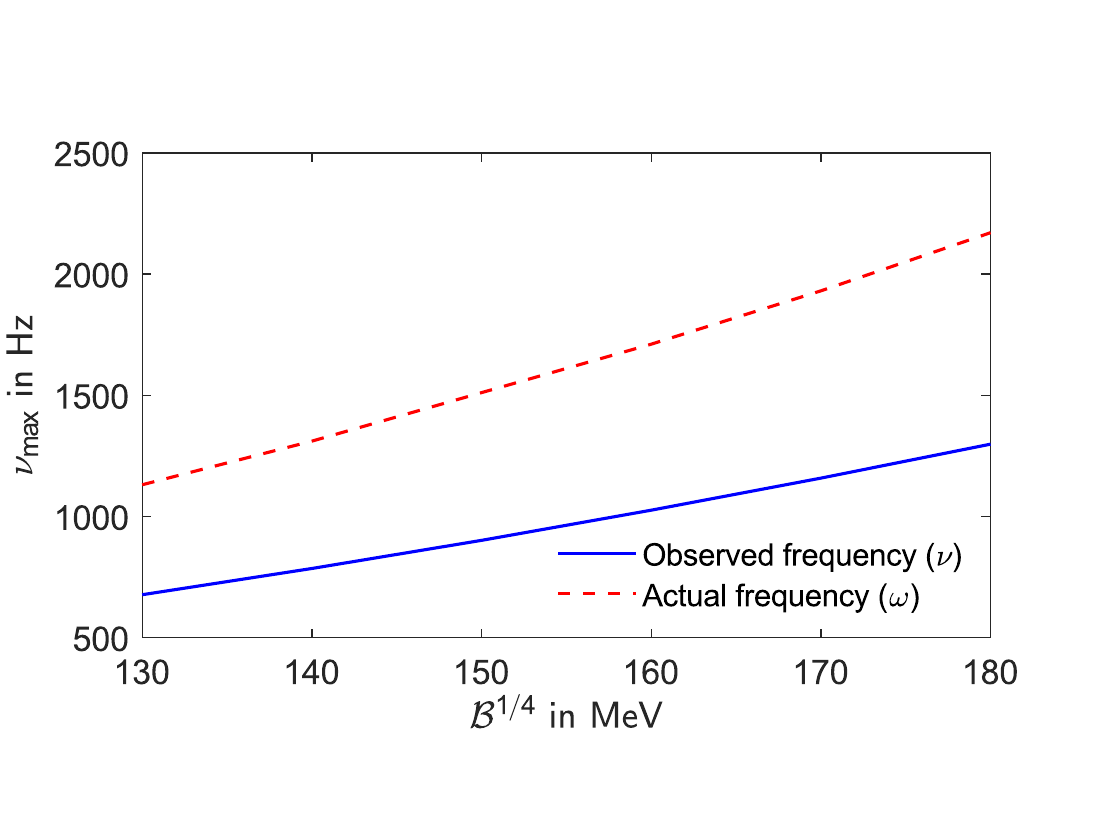}
			\caption{\label{fig:nu_max bag-1} Limiting observed frequency ($\nu_{\text{max}}$) and corresponding actual frequency $\omega$ vs $\mathcal{B}^{1/4}$ graph.}
		\end{figure}

		The limiting frequencies ($\nu_{\text{max}}$) are found to be different for different values of bag constant. The corresponding variation of $\nu_{\text{max}}$ is plotted in Fig.~\ref{fig:nu_max bag-1}, showing that, the $\nu_{\text{max}}$ increases almost linearly with $\mathcal{B}^{1/4}$.
		Several numerical models and simulations \cite{bag145,lim_bag, bag_sp1} indicate that, the value of the bag constant for stable quark matter system lies within the range of $\sim (130~\rm{MeV})^4-(162~\rm{MeV})^4$. This range of bag constant corresponds to the observed frequency ($\nu$) range $677.4\sim 1052.6$ Hz (see Fig.~\ref{fig:nu_max bag-1}). As a consequence, the span of $677.4\sim 1052.6$ Hz addresses the possible upper-bound of the observed frequency ($\nu$) of quark stars (millisecond order) \cite{atnf}. In contrast, these values of bag constant ($(130~\rm{MeV})^4 \sim (162~\rm{MeV})^4$) indicate the range on actual frequency ($\omega$) $1131 \sim 1755.0$ Hz, which agrees with the work of \citet{upperlim_fr}. For example, according to the best fit point of our $\chi^2$ analysis ($\mathcal{B}=(136.5~\rm{MeV})^4$), the maximum value of observed frequency ($\nu_{\text{max}}$) of quark star is $\sim 747.7346$ Hz (agrees with observational evidences of fast spinning pulsar). However, the corresponding actual frequency ($\omega_{\text{max}}$) for that case is $\sim 1248$ Hz.
		
		
		In the current work, we have looked at a significant quantity given by the  radius to the Schwarzschild radius ratio ($\frac{R_{\text{max}}}{R_{\rm{sch}}}$, inverse of compactness) of the star. The variation of the ratio ($\frac{R_{\text{max}}}{R_{\rm{sch}}}$) with frequency ($\nu$) are traced out for different values of bag constants (see Fig.~\ref{fig:max nu}c). In the low frequency range ($\omega<200$ Hz), the numerical value of the ratio is found to be independent of the bag constant and remains almost unchanged ($\sim 2.2903$ in the geometric unit, as obtained from the work of \cite{SBa}) with frequency (Fig.~\ref{fig:max nu}c). But it starts falling gradually with increasing frequency as the rotational frequency reaches $300$ Hz. In this range of frequencies, a minute dependence of the bag constant is observable. However, at higher frequencies, all the $\frac{R_{\text{max}}}{R_{\rm{sch}}}$ curves for the different bag constants suffer a quick fall toward unity. From the above study, it is evident that the compactness of the quark star is extremely high, but it cannot become a black hole even in the limiting case. Thus they (strange stars) and black hole could co-exist as candidates of cold dark matter. While their phenomenological signatures may not be sufficient to distinguish between them, their different signatures in the gravitational wave scenario may be interesting to study \cite{ssrmrp}.

	\section{Conclusion \label{sec_conc}}
		In this work, we have studied the limiting mass for the rotating quark star, which is introduced here as the ``Chandrasekhar limit for rotating quark stars". The limit mostly depends on the universal constants and the Bag parameter as well as the angular velocity of the star. The maximum possible observed frequency ($\nu_{\max}$) is also estimated in the current work, which agrees with recent observations \cite{J0740+6620,J1614-2230,B1957+20,J1600-3053,J2215+5135,J0751+1807,J1311-3430,atnf} as well as results based on numerical simulation \cite{upperlim_fr}. A relation between the limiting radius ($R_{\text{max}}$) and corresponding Schwarzschild radius ($R_{\rm{sch}}$) for a range of rotational frequencies has also been resolved in our current work. The numerical value of the quantity $\frac{R_{\text{max}}}{R_{\rm{sch}}}$ lie at $\sim 2.2903$ for a static star (close to the work of \citet{SBa} (i.e. 2.6667), where the relativistic correction in volume is not taken into account), however it tends toward unity for an extreme case of the rotation($\omega \rightarrow R_{\text{max}} c$). But for a fast spinning star, the Schwarzschild radius cannot address the event horizon \cite{hobson}. So, in order to obtain the event horizon for such spinning bodies, the Kerr space time has to be taken into account, which provides a comparatively smaller horizon than that of the Schwarzschild metric system \cite{hobson}. Consequently, one can conclude that a quark star can never behave as black hole.


	\section*{Acknowledgment}	
		Two of the authors (S.B. and A.H.) wish to acknowledge the support received from St. Xavier's College. A.H. also acknowledges the University Grant Commission (UGC) of the Government of India, for providing financial support, in the form of UGC-CSIR NET-JRF. The work of S.R. was performed under the aegis of the Raja Ramanna Fellowship of the Department of Atomic Energy, Govt. of India. S.R. also gratefully acknowledges the hospitality of Frankfurt Institute of Advance Studies (FIAS).

	\bibliography{Project1}

\begin{thebibliography}{60}%
\makeatletter
\providecommand \@ifxundefined [1]{%
 \@ifx{#1\undefined}
}%
\providecommand \@ifnum [1]{%
 \ifnum #1\expandafter \@firstoftwo
 \else \expandafter \@secondoftwo
 \fi
}%
\providecommand \@ifx [1]{%
 \ifx #1\expandafter \@firstoftwo
 \else \expandafter \@secondoftwo
 \fi
}%
\providecommand \natexlab [1]{#1}%
\providecommand \enquote  [1]{``#1''}%
\providecommand \bibnamefont  [1]{#1}%
\providecommand \bibfnamefont [1]{#1}%
\providecommand \citenamefont [1]{#1}%
\providecommand \href@noop [0]{\@secondoftwo}%
\providecommand \href [0]{\begingroup \@sanitize@url \@href}%
\providecommand \@href[1]{\@@startlink{#1}\@@href}%
\providecommand \@@href[1]{\endgroup#1\@@endlink}%
\providecommand \@sanitize@url [0]{\catcode `\\12\catcode `\$12\catcode
  `\&12\catcode `\#12\catcode `\^12\catcode `\_12\catcode `\%12\relax}%
\providecommand \@@startlink[1]{}%
\providecommand \@@endlink[0]{}%
\providecommand \url  [0]{\begingroup\@sanitize@url \@url }%
\providecommand \@url [1]{\endgroup\@href {#1}{\urlprefix }}%
\providecommand \urlprefix  [0]{URL }%
\providecommand \Eprint [0]{\href }%
\providecommand \doibase [0]{http://dx.doi.org/}%
\providecommand \selectlanguage [0]{\@gobble}%
\providecommand \bibinfo  [0]{\@secondoftwo}%
\providecommand \bibfield  [0]{\@secondoftwo}%
\providecommand \translation [1]{[#1]}%
\providecommand \BibitemOpen [0]{}%
\providecommand \bibitemStop [0]{}%
\providecommand \bibitemNoStop [0]{.\EOS\space}%
\providecommand \EOS [0]{\spacefactor3000\relax}%
\providecommand \BibitemShut  [1]{\csname bibitem#1\endcsname}%
\let\auto@bib@innerbib\@empty
\bibitem [{\citenamefont {{Ivanenko}}\ and\ \citenamefont
  {{Kurdgelaidze}}(1965)}]{ivan_1}%
  \BibitemOpen
  \bibfield  {author} {\bibinfo {author} {\bibfnamefont {D.~D.}\ \bibnamefont
  {{Ivanenko}}}\ and\ \bibinfo {author} {\bibfnamefont {D.~F.}\ \bibnamefont
  {{Kurdgelaidze}}},\ }\href {\doibase 10.1007/BF01042830} {\bibfield
  {journal} {\bibinfo  {journal} {Astrophysics}\ }\textbf {\bibinfo {volume}
  {1}},\ \bibinfo {pages} {251} (\bibinfo {year} {1965})}\BibitemShut {NoStop}%
\bibitem [{\citenamefont {{Ivanenko}}\ and\ \citenamefont
  {{Kurdgelaidze}}(1969)}]{ivan_2}%
  \BibitemOpen
  \bibfield  {author} {\bibinfo {author} {\bibfnamefont {D.}~\bibnamefont
  {{Ivanenko}}}\ and\ \bibinfo {author} {\bibfnamefont {D.~F.}\ \bibnamefont
  {{Kurdgelaidze}}},\ }\href {\doibase 10.1007/BF02753988} {\bibfield
  {journal} {\bibinfo  {journal} {Nuovo Cimento Lettere}\ }\textbf {\bibinfo
  {volume} {2}},\ \bibinfo {pages} {13} (\bibinfo {year} {1969})}\BibitemShut
  {NoStop}%
\bibitem [{\citenamefont {Griffiths}(2008)}]{Quark_Wikipedia}%
  \BibitemOpen
  \bibfield  {author} {\bibinfo {author} {\bibfnamefont {D.}~\bibnamefont
  {Griffiths}},\ }\href@noop {} {\emph {\bibinfo {title} {Introduction to
  elementary particles}}}\ (\bibinfo  {publisher} {John Wiley \& Sons},\
  \bibinfo {year} {2008})\BibitemShut {NoStop}%
\bibitem [{\citenamefont {Yu}\ and\ \citenamefont {Xu}(2010)}]{hyu}%
  \BibitemOpen
  \bibfield  {author} {\bibinfo {author} {\bibfnamefont {H.~W.}\ \bibnamefont
  {Yu}}\ and\ \bibinfo {author} {\bibfnamefont {R.~X.}\ \bibnamefont {Xu}},\
  }\href@noop {} {\bibfield  {journal} {\bibinfo  {journal} {Res. Astron.
  Astrophys.}\ }\textbf {\bibinfo {volume} {11}},\ \bibinfo {pages} {471}
  (\bibinfo {year} {2010})}\BibitemShut {NoStop}%
\bibitem [{\citenamefont {Prakash}\ \emph {et~al.}(2003)\citenamefont
  {Prakash}, \citenamefont {Lattimer}, \citenamefont {Steiner},\ and\
  \citenamefont {Page}}]{Prakash:2002xx}%
  \BibitemOpen
  \bibfield  {author} {\bibinfo {author} {\bibfnamefont {M.}~\bibnamefont
  {Prakash}}, \bibinfo {author} {\bibfnamefont {J.~M.}\ \bibnamefont
  {Lattimer}}, \bibinfo {author} {\bibfnamefont {A.~W.}\ \bibnamefont
  {Steiner}}, \ and\ \bibinfo {author} {\bibfnamefont {D.}~\bibnamefont
  {Page}},\ }\bibfield  {booktitle} {\emph {\bibinfo {booktitle} {{Proceedings,
  16th International Conference on Ultra-Relativistic nucleus nucleus
  collisions (Quark Matter 2012): Nantes, France, July 18-24, 2002}}},\ }\href
  {\doibase 10.1016/S0375-9474(02)01514-2} {\bibfield  {journal} {\bibinfo
  {journal} {Nucl. Phys.}\ }\textbf {\bibinfo {volume} {A715}},\ \bibinfo
  {pages} {835} (\bibinfo {year} {2003})}\BibitemShut {NoStop}%
\bibitem [{\citenamefont {Jaikumar}\ \emph {et~al.}(2004)\citenamefont
  {Jaikumar}, \citenamefont {Gale}, \citenamefont {Page},\ and\ \citenamefont
  {Prakash}}]{Jaikumar:2004zy}%
  \BibitemOpen
  \bibfield  {author} {\bibinfo {author} {\bibfnamefont {P.}~\bibnamefont
  {Jaikumar}}, \bibinfo {author} {\bibfnamefont {C.}~\bibnamefont {Gale}},
  \bibinfo {author} {\bibfnamefont {D.}~\bibnamefont {Page}}, \ and\ \bibinfo
  {author} {\bibfnamefont {M.}~\bibnamefont {Prakash}},\ }\bibfield
  {booktitle} {\emph {\bibinfo {booktitle} {{High energy physics. Proceedings,
  26th Annual Montreal-Rochester-Syracuse-Toronto Conference, MRST 2004,
  Montreal, Canada, May 12-14, 2004}}},\ }\href {\doibase
  10.1142/S0217751X04022566} {\bibfield  {journal} {\bibinfo  {journal} {Int.
  J. Mod. Phys.}\ }\textbf {\bibinfo {volume} {A19}},\ \bibinfo {pages} {5335}
  (\bibinfo {year} {2004})}\BibitemShut {NoStop}%
\bibitem [{\citenamefont {Witten}(1984)}]{Witten}%
  \BibitemOpen
  \bibfield  {author} {\bibinfo {author} {\bibfnamefont {E.}~\bibnamefont
  {Witten}},\ }\href {\doibase 10.1103/PhysRevD.30.272} {\bibfield  {journal}
  {\bibinfo  {journal} {Phys. Rev. D}\ }\textbf {\bibinfo {volume} {30}},\
  \bibinfo {pages} {272} (\bibinfo {year} {1984})}\BibitemShut {NoStop}%
\bibitem [{\citenamefont {Li}\ \emph {et~al.}(2010)\citenamefont {Li},
  \citenamefont {Luo},\ and\ \citenamefont {Zong}}]{Bag_Model_2}%
  \BibitemOpen
  \bibfield  {author} {\bibinfo {author} {\bibfnamefont {H.}~\bibnamefont
  {Li}}, \bibinfo {author} {\bibfnamefont {X.-L.}\ \bibnamefont {Luo}}, \ and\
  \bibinfo {author} {\bibfnamefont {H.-S.}\ \bibnamefont {Zong}},\ }\href
  {\doibase 10.1103/PhysRevD.82.065017} {\bibfield  {journal} {\bibinfo
  {journal} {Phys. Rev. D}\ }\textbf {\bibinfo {volume} {82}},\ \bibinfo
  {pages} {065017} (\bibinfo {year} {2010})}\BibitemShut {NoStop}%
\bibitem [{\citenamefont {Bhattacharyya}\ \emph {et~al.}(2006)\citenamefont
  {Bhattacharyya}, \citenamefont {Ghosh}, \citenamefont {Joarder},
  \citenamefont {Mallick},\ and\ \citenamefont {Raha}}]{ns_to_qs}%
  \BibitemOpen
  \bibfield  {author} {\bibinfo {author} {\bibfnamefont {A.}~\bibnamefont
  {Bhattacharyya}}, \bibinfo {author} {\bibfnamefont {S.~K.}\ \bibnamefont
  {Ghosh}}, \bibinfo {author} {\bibfnamefont {P.~S.}\ \bibnamefont {Joarder}},
  \bibinfo {author} {\bibfnamefont {R.}~\bibnamefont {Mallick}}, \ and\
  \bibinfo {author} {\bibfnamefont {S.}~\bibnamefont {Raha}},\ }\href {\doibase
  10.1103/PhysRevC.74.065804} {\bibfield  {journal} {\bibinfo  {journal} {Phys.
  Rev. C}\ }\textbf {\bibinfo {volume} {74}},\ \bibinfo {pages} {065804}
  (\bibinfo {year} {2006})}\BibitemShut {NoStop}%
\bibitem [{\citenamefont {{Alcock}}\ \emph {et~al.}(1986)\citenamefont
  {{Alcock}}, \citenamefont {{Farhi}},\ and\ \citenamefont
  {{Olinto}}}]{alcock}%
  \BibitemOpen
  \bibfield  {author} {\bibinfo {author} {\bibfnamefont {C.}~\bibnamefont
  {{Alcock}}}, \bibinfo {author} {\bibfnamefont {E.}~\bibnamefont {{Farhi}}}, \
  and\ \bibinfo {author} {\bibfnamefont {A.}~\bibnamefont {{Olinto}}},\ }\href
  {\doibase 10.1086/164679} {\bibfield  {journal} {\bibinfo  {journal} {\apj}\
  }\textbf {\bibinfo {volume} {310}},\ \bibinfo {pages} {261} (\bibinfo {year}
  {1986})}\BibitemShut {NoStop}%
\bibitem [{\citenamefont {Rohlf}(1994)}]{Bag_Model_1}%
  \BibitemOpen
  \bibfield  {author} {\bibinfo {author} {\bibfnamefont {J.}~\bibnamefont
  {Rohlf}},\ }\href@noop {} {\emph {\bibinfo {title} {Modern Physics from alpha
  to Z0}}}\ (\bibinfo  {publisher} {Wiley},\ \bibinfo {year}
  {1994})\BibitemShut {NoStop}%
\bibitem [{\citenamefont {{Haensel}}\ \emph {et~al.}(1986)\citenamefont
  {{Haensel}}, \citenamefont {{Zdunik}},\ and\ \citenamefont
  {{Schaefer}}}]{Strange_qs_1}%
  \BibitemOpen
  \bibfield  {author} {\bibinfo {author} {\bibfnamefont {P.}~\bibnamefont
  {{Haensel}}}, \bibinfo {author} {\bibfnamefont {J.~L.}\ \bibnamefont
  {{Zdunik}}}, \ and\ \bibinfo {author} {\bibfnamefont {R.}~\bibnamefont
  {{Schaefer}}},\ }\href@noop {} {\bibfield  {journal} {\bibinfo  {journal}
  {A\&A}\ }\textbf {\bibinfo {volume} {160}},\ \bibinfo {pages} {121} (\bibinfo
  {year} {1986})}\BibitemShut {NoStop}%
\bibitem [{\citenamefont {Ghosh}\ and\ \citenamefont
  {Sahu}(1993)}]{Strange_qs_2}%
  \BibitemOpen
  \bibfield  {author} {\bibinfo {author} {\bibfnamefont {S.~K.}\ \bibnamefont
  {Ghosh}}\ and\ \bibinfo {author} {\bibfnamefont {P.~K.}\ \bibnamefont
  {Sahu}},\ }\href@noop {} {\bibfield  {journal} {\bibinfo  {journal}
  {International Journal of Modern Physics E}\ }\textbf {\bibinfo {volume}
  {2}},\ \bibinfo {pages} {575} (\bibinfo {year} {1993})}\BibitemShut {NoStop}%
\bibitem [{\citenamefont {Weber}\ \emph {et~al.}(2012)\citenamefont {Weber},
  \citenamefont {Orsaria}, \citenamefont {Rodrigues},\ and\ \citenamefont
  {Yang}}]{weber_structure}%
  \BibitemOpen
  \bibfield  {author} {\bibinfo {author} {\bibfnamefont {F.}~\bibnamefont
  {Weber}}, \bibinfo {author} {\bibfnamefont {M.}~\bibnamefont {Orsaria}},
  \bibinfo {author} {\bibfnamefont {H.}~\bibnamefont {Rodrigues}}, \ and\
  \bibinfo {author} {\bibfnamefont {S.-H.}\ \bibnamefont {Yang}},\ }\href
  {\doibase 10.1017/S1743921312023174} {\bibfield  {journal} {\bibinfo
  {journal} {Proceedings of the International Astronomical Union}\ }\textbf
  {\bibinfo {volume} {8}} (\bibinfo {year} {2012}),\
  10.1017/S1743921312023174}\BibitemShut {NoStop}%
\bibitem [{\citenamefont {Madsen}(1999)}]{Strange_Q_Star}%
  \BibitemOpen
  \bibfield  {author} {\bibinfo {author} {\bibfnamefont {J.}~\bibnamefont
  {Madsen}},\ }\bibfield  {booktitle} {\emph {\bibinfo {booktitle} {{Hadrons in
  dense matter and hadrosynthesis. Proceedings, 11th Chris Engelbrecht Summer
  School, Cape Town, South Africa, February 4-13, 1998}}},\ }\href {\doibase
  10.1007/BFb0107314} {\bibfield  {journal} {\bibinfo  {journal} {Lect. Notes
  Phys.}\ }\textbf {\bibinfo {volume} {516}},\ \bibinfo {pages} {162} (\bibinfo
  {year} {1999})}\BibitemShut {NoStop}%
\bibitem [{\citenamefont {Weber}(2001)}]{Pure_qs}%
  \BibitemOpen
  \bibfield  {author} {\bibinfo {author} {\bibfnamefont {F.}~\bibnamefont
  {Weber}},\ }\href@noop {} {\bibfield  {journal} {\bibinfo  {journal} {Journal
  of Physics G: Nuclear and Particle Physics}\ }\textbf {\bibinfo {volume}
  {27}},\ \bibinfo {pages} {465} (\bibinfo {year} {2001})}\BibitemShut
  {NoStop}%
\bibitem [{\citenamefont {Banerjee}\ \emph {et~al.}(2003)\citenamefont
  {Banerjee}, \citenamefont {Bhattacharyya}, \citenamefont {Ghosh},
  \citenamefont {Raha}, \citenamefont {Sinha},\ and\ \citenamefont
  {Toki}}]{Quark_galaxy}%
  \BibitemOpen
  \bibfield  {author} {\bibinfo {author} {\bibfnamefont {S.}~\bibnamefont
  {Banerjee}}, \bibinfo {author} {\bibfnamefont {A.}~\bibnamefont
  {Bhattacharyya}}, \bibinfo {author} {\bibfnamefont {S.~K.}\ \bibnamefont
  {Ghosh}}, \bibinfo {author} {\bibfnamefont {S.}~\bibnamefont {Raha}},
  \bibinfo {author} {\bibfnamefont {B.}~\bibnamefont {Sinha}}, \ and\ \bibinfo
  {author} {\bibfnamefont {H.}~\bibnamefont {Toki}},\ }\href@noop {} {\bibfield
   {journal} {\bibinfo  {journal} {Monthly Notices of the Royal Astronomical
  Society}\ }\textbf {\bibinfo {volume} {340}},\ \bibinfo {pages} {284}
  (\bibinfo {year} {2003})}\BibitemShut {NoStop}%
\bibitem [{\citenamefont {Fredriksson}\ \emph {et~al.}(1998)\citenamefont
  {Fredriksson}, \citenamefont {Enstrom}, \citenamefont {Hansson},
  \citenamefont {Ekelin},\ and\ \citenamefont {Nicolaidis}}]{qg1}%
  \BibitemOpen
  \bibfield  {author} {\bibinfo {author} {\bibfnamefont {S.}~\bibnamefont
  {Fredriksson}}, \bibinfo {author} {\bibfnamefont {D.}~\bibnamefont
  {Enstrom}}, \bibinfo {author} {\bibfnamefont {J.}~\bibnamefont {Hansson}},
  \bibinfo {author} {\bibfnamefont {S.}~\bibnamefont {Ekelin}}, \ and\ \bibinfo
  {author} {\bibfnamefont {A.}~\bibnamefont {Nicolaidis}},\ }in\ \href@noop {}
  {\emph {\bibinfo {booktitle} {{Proceedings, 2nd International Heidelberg
  Conference on Dark matter in astrophysics and particle physics (DARK 1998):
  Heidelberg, Germany, July 20-25, 1998}}}}\ (\bibinfo {year} {1998})\ pp.\
  \bibinfo {pages} {651--658},\ \Eprint {http://arxiv.org/abs/astro-ph/9810389}
  {arXiv:astro-ph/9810389 [astro-ph]} \BibitemShut {NoStop}%
\bibitem [{\citenamefont {Enstrom}\ \emph {et~al.}(1998)\citenamefont
  {Enstrom}, \citenamefont {Fredriksson}, \citenamefont {Hansson},
  \citenamefont {Nicolaidis},\ and\ \citenamefont {Ekelin}}]{qg2}%
  \BibitemOpen
  \bibfield  {author} {\bibinfo {author} {\bibfnamefont {D.}~\bibnamefont
  {Enstrom}}, \bibinfo {author} {\bibfnamefont {S.}~\bibnamefont
  {Fredriksson}}, \bibinfo {author} {\bibfnamefont {J.}~\bibnamefont
  {Hansson}}, \bibinfo {author} {\bibfnamefont {A.}~\bibnamefont {Nicolaidis}},
  \ and\ \bibinfo {author} {\bibfnamefont {S.}~\bibnamefont {Ekelin}},\
  }\href@noop {} {\  (\bibinfo {year} {1998})},\ \Eprint
  {http://arxiv.org/abs/astro-ph/9802236} {arXiv:astro-ph/9802236 [astro-ph]}
  \BibitemShut {NoStop}%
\bibitem [{\citenamefont {{Overgard}}\ and\ \citenamefont
  {{Ostgaard}}(1991)}]{tov}%
  \BibitemOpen
  \bibfield  {author} {\bibinfo {author} {\bibfnamefont {T.}~\bibnamefont
  {{Overgard}}}\ and\ \bibinfo {author} {\bibfnamefont {E.}~\bibnamefont
  {{Ostgaard}}},\ }\href@noop {} {\bibfield  {journal} {\bibinfo  {journal}
  {A\&A}\ }\textbf {\bibinfo {volume} {243}},\ \bibinfo {pages} {412} (\bibinfo
  {year} {1991})}\BibitemShut {NoStop}%
\bibitem [{\citenamefont {Burgio}\ \emph {et~al.}(2002)\citenamefont {Burgio},
  \citenamefont {Baldo}, \citenamefont {Sahu}, \citenamefont {Santra},\ and\
  \citenamefont {Schulze}}]{lm1}%
  \BibitemOpen
  \bibfield  {author} {\bibinfo {author} {\bibfnamefont {G.}~\bibnamefont
  {Burgio}}, \bibinfo {author} {\bibfnamefont {M.}~\bibnamefont {Baldo}},
  \bibinfo {author} {\bibfnamefont {P.}~\bibnamefont {Sahu}}, \bibinfo {author}
  {\bibfnamefont {A.}~\bibnamefont {Santra}}, \ and\ \bibinfo {author}
  {\bibfnamefont {H.-J.}\ \bibnamefont {Schulze}},\ }\href {\doibase
  https://doi.org/10.1016/S0370-2693(01)01479-4} {\bibfield  {journal}
  {\bibinfo  {journal} {Physics Letters B}\ }\textbf {\bibinfo {volume}
  {526}},\ \bibinfo {pages} {19 } (\bibinfo {year} {2002})}\BibitemShut
  {NoStop}%
\bibitem [{\citenamefont {{Harko, T.}}\ and\ \citenamefont {{Cheng, K.
  S.}}(2002)}]{lm2}%
  \BibitemOpen
  \bibfield  {author} {\bibinfo {author} {\bibnamefont {{Harko, T.}}}\ and\
  \bibinfo {author} {\bibnamefont {{Cheng, K. S.}}},\ }\href {\doibase
  10.1051/0004-6361:20020260} {\bibfield  {journal} {\bibinfo  {journal}
  {A\&A}\ }\textbf {\bibinfo {volume} {385}},\ \bibinfo {pages} {947} (\bibinfo
  {year} {2002})}\BibitemShut {NoStop}%
\bibitem [{\citenamefont {Bozzola}\ \emph {et~al.}(2019)\citenamefont
  {Bozzola}, \citenamefont {Espino}, \citenamefont {Lewin},\ and\ \citenamefont
  {Paschalidis}}]{rotlm}%
  \BibitemOpen
  \bibfield  {author} {\bibinfo {author} {\bibfnamefont {G.}~\bibnamefont
  {Bozzola}}, \bibinfo {author} {\bibfnamefont {P.~L.}\ \bibnamefont {Espino}},
  \bibinfo {author} {\bibfnamefont {C.~D.}\ \bibnamefont {Lewin}}, \ and\
  \bibinfo {author} {\bibfnamefont {V.}~\bibnamefont {Paschalidis}},\ }\href
  {\doibase 10.1140/epja/i2019-12831-2} {\bibfield  {journal} {\bibinfo
  {journal} {Eur. Phys. J.}\ }\textbf {\bibinfo {volume} {A55}},\ \bibinfo
  {pages} {149} (\bibinfo {year} {2019})}\BibitemShut {NoStop}%
\bibitem [{\citenamefont {Szkudlarek}\ \emph {et~al.}(2019)\citenamefont
  {Szkudlarek}, \citenamefont {Gondek-Rosinska}, \citenamefont {Villain},\ and\
  \citenamefont {Ansorg}}]{rotlm_2}%
  \BibitemOpen
  \bibfield  {author} {\bibinfo {author} {\bibfnamefont {M.}~\bibnamefont
  {Szkudlarek}}, \bibinfo {author} {\bibfnamefont {D.}~\bibnamefont
  {Gondek-Rosinska}}, \bibinfo {author} {\bibfnamefont {L.}~\bibnamefont
  {Villain}}, \ and\ \bibinfo {author} {\bibfnamefont {M.}~\bibnamefont
  {Ansorg}},\ }\href {\doibase 10.3847/1538-4357/ab1752} {\bibfield  {journal}
  {\bibinfo  {journal} {Astrophys. J.}\ }\textbf {\bibinfo {volume} {879}},\
  \bibinfo {pages} {44} (\bibinfo {year} {2019})}\BibitemShut {NoStop}%
\bibitem [{\citenamefont {Shapiro}\ and\ \citenamefont
  {Teukolsky}(1983)}]{shapiro}%
  \BibitemOpen
  \bibfield  {author} {\bibinfo {author} {\bibfnamefont {S.~L.}\ \bibnamefont
  {Shapiro}}\ and\ \bibinfo {author} {\bibfnamefont {S.~A.}\ \bibnamefont
  {Teukolsky}},\ }\href@noop {} {\emph {\bibinfo {title} {{Black holes, white
  dwarfs, and neutron stars: The physics of compact objects}}}}\ (\bibinfo
  {publisher} {Wiley \& Sons},\ \bibinfo {year} {1983})\BibitemShut {NoStop}%
\bibitem [{\citenamefont {Zhou}(2017)}]{def_agnst}%
  \BibitemOpen
  \bibfield  {author} {\bibinfo {author} {\bibfnamefont {E.}~\bibnamefont
  {Zhou}},\ }\href {\doibase 10.1088/1742-6596/861/1/012007} {\bibfield
  {journal} {\bibinfo  {journal} {Journal of Physics: Conference Series}\
  }\textbf {\bibinfo {volume} {861}},\ \bibinfo {pages} {012007} (\bibinfo
  {year} {2017})}\BibitemShut {NoStop}%
\bibitem [{\citenamefont {Banerjee}\ \emph {et~al.}(2000)\citenamefont
  {Banerjee}, \citenamefont {Ghosh},\ and\ \citenamefont {Raha}}]{SBa}%
  \BibitemOpen
  \bibfield  {author} {\bibinfo {author} {\bibfnamefont {S.}~\bibnamefont
  {Banerjee}}, \bibinfo {author} {\bibfnamefont {S.~K.}\ \bibnamefont {Ghosh}},
  \ and\ \bibinfo {author} {\bibfnamefont {S.}~\bibnamefont {Raha}},\
  }\href@noop {} {\bibfield  {journal} {\bibinfo  {journal} {Journal of Physics
  G: Nuclear and Particle Physics}\ }\textbf {\bibinfo {volume} {26}},\
  \bibinfo {pages} {L1} (\bibinfo {year} {2000})}\BibitemShut {NoStop}%
\bibitem [{Lan(1965)}]{Landau}%
  \BibitemOpen
  in\ \href {\doibase https://doi.org/10.1016/B978-0-08-010586-4.50013-4}
  {\emph {\bibinfo {booktitle} {Collected Papers of L.D. Landau}}},\ \bibinfo
  {editor} {edited by\ \bibinfo {editor} {\bibfnamefont {D.~T.}\ \bibnamefont
  {HAAR}}}\ (\bibinfo  {publisher} {Pergamon},\ \bibinfo {year} {1965})\ pp.\
  \bibinfo {pages} {60 -- 62}\BibitemShut {NoStop}%
\bibitem [{\citenamefont {Yazdizadeh}\ and\ \citenamefont
  {Bordbar}(2013)}]{Bag_Density}%
  \BibitemOpen
  \bibfield  {author} {\bibinfo {author} {\bibfnamefont {T.}~\bibnamefont
  {Yazdizadeh}}\ and\ \bibinfo {author} {\bibfnamefont {G.~H.}\ \bibnamefont
  {Bordbar}},\ }\href {\doibase 10.1007/s10511-013-9272-y} {\bibfield
  {journal} {\bibinfo  {journal} {Astrophysics}\ }\textbf {\bibinfo {volume}
  {56}},\ \bibinfo {pages} {121} (\bibinfo {year} {2013})}\BibitemShut
  {NoStop}%
\bibitem [{\citenamefont {Owen}(2005)}]{def_sphere_1}%
  \BibitemOpen
  \bibfield  {author} {\bibinfo {author} {\bibfnamefont {B.~J.}\ \bibnamefont
  {Owen}},\ }\href {\doibase 10.1103/PhysRevLett.95.211101} {\bibfield
  {journal} {\bibinfo  {journal} {Phys. Rev. Lett.}\ }\textbf {\bibinfo
  {volume} {95}},\ \bibinfo {pages} {211101} (\bibinfo {year}
  {2005})}\BibitemShut {NoStop}%
\bibitem [{\citenamefont {Ushomirsky}\ \emph {et~al.}(2000)\citenamefont
  {Ushomirsky}, \citenamefont {Cutler},\ and\ \citenamefont
  {Bildsten}}]{ushomirsky2000}%
  \BibitemOpen
  \bibfield  {author} {\bibinfo {author} {\bibfnamefont {G.}~\bibnamefont
  {Ushomirsky}}, \bibinfo {author} {\bibfnamefont {C.}~\bibnamefont {Cutler}},
  \ and\ \bibinfo {author} {\bibfnamefont {L.}~\bibnamefont {Bildsten}},\
  }\href@noop {} {\bibfield  {journal} {\bibinfo  {journal} {Monthly Notices of
  the Royal Astronomical Society}\ }\textbf {\bibinfo {volume} {319}},\
  \bibinfo {pages} {902} (\bibinfo {year} {2000})}\BibitemShut {NoStop}%
\bibitem [{\citenamefont {DeGrand}\ \emph {et~al.}(1975)\citenamefont
  {DeGrand}, \citenamefont {Jaffe}, \citenamefont {Johnson},\ and\
  \citenamefont {Kiskis}}]{bag_sp1}%
  \BibitemOpen
  \bibfield  {author} {\bibinfo {author} {\bibfnamefont {T.}~\bibnamefont
  {DeGrand}}, \bibinfo {author} {\bibfnamefont {R.~L.}\ \bibnamefont {Jaffe}},
  \bibinfo {author} {\bibfnamefont {K.}~\bibnamefont {Johnson}}, \ and\
  \bibinfo {author} {\bibfnamefont {J.}~\bibnamefont {Kiskis}},\ }\href
  {\doibase 10.1103/PhysRevD.12.2060} {\bibfield  {journal} {\bibinfo
  {journal} {Phys. Rev. D}\ }\textbf {\bibinfo {volume} {12}},\ \bibinfo
  {pages} {2060} (\bibinfo {year} {1975})}\BibitemShut {NoStop}%
\bibitem [{\citenamefont {Chodos}\ \emph
  {et~al.}(1974{\natexlab{a}})\citenamefont {Chodos}, \citenamefont {Jaffe},
  \citenamefont {Johnson},\ and\ \citenamefont {Thorn}}]{bag_sp2}%
  \BibitemOpen
  \bibfield  {author} {\bibinfo {author} {\bibfnamefont {A.}~\bibnamefont
  {Chodos}}, \bibinfo {author} {\bibfnamefont {R.~L.}\ \bibnamefont {Jaffe}},
  \bibinfo {author} {\bibfnamefont {K.}~\bibnamefont {Johnson}}, \ and\
  \bibinfo {author} {\bibfnamefont {C.~B.}\ \bibnamefont {Thorn}},\ }\href
  {\doibase 10.1103/PhysRevD.10.2599} {\bibfield  {journal} {\bibinfo
  {journal} {Phys. Rev. D}\ }\textbf {\bibinfo {volume} {10}},\ \bibinfo
  {pages} {2599} (\bibinfo {year} {1974}{\natexlab{a}})}\BibitemShut {NoStop}%
\bibitem [{\citenamefont {Chodos}\ \emph
  {et~al.}(1974{\natexlab{b}})\citenamefont {Chodos}, \citenamefont {Jaffe},
  \citenamefont {Johnson}, \citenamefont {Thorn},\ and\ \citenamefont
  {Weisskopf}}]{bag_sp3}%
  \BibitemOpen
  \bibfield  {author} {\bibinfo {author} {\bibfnamefont {A.}~\bibnamefont
  {Chodos}}, \bibinfo {author} {\bibfnamefont {R.~L.}\ \bibnamefont {Jaffe}},
  \bibinfo {author} {\bibfnamefont {K.}~\bibnamefont {Johnson}}, \bibinfo
  {author} {\bibfnamefont {C.~B.}\ \bibnamefont {Thorn}}, \ and\ \bibinfo
  {author} {\bibfnamefont {V.~F.}\ \bibnamefont {Weisskopf}},\ }\href {\doibase
  10.1103/PhysRevD.9.3471} {\bibfield  {journal} {\bibinfo  {journal} {Phys.
  Rev. D}\ }\textbf {\bibinfo {volume} {9}},\ \bibinfo {pages} {3471} (\bibinfo
  {year} {1974}{\natexlab{b}})}\BibitemShut {NoStop}%
\bibitem [{\citenamefont {Alaverdyan}\ and\ \citenamefont
  {Vartanyan}(2017)}]{mmhs}%
  \BibitemOpen
  \bibfield  {author} {\bibinfo {author} {\bibfnamefont {G.}~\bibnamefont
  {Alaverdyan}}\ and\ \bibinfo {author} {\bibfnamefont {Y.}~\bibnamefont
  {Vartanyan}},\ }\href {\doibase 10.1007/s10511-017-9507-4} {\bibfield
  {journal} {\bibinfo  {journal} {Astrophysics}\ }\textbf {\bibinfo {volume}
  {60}} (\bibinfo {year} {2017}),\ 10.1007/s10511-017-9507-4}\BibitemShut
  {NoStop}%
\bibitem [{\citenamefont {C.~Kittel}(1969)}]{Fermi_Hyperphysics}%
  \BibitemOpen
  \bibfield  {author} {\bibinfo {author} {\bibfnamefont {K.~H.}\ \bibnamefont
  {C.~Kittel}},\ }\href@noop {} {\emph {\bibinfo {title} {Thermal Physics (2nd
  Edition)}}}\ (\bibinfo  {publisher} {W. H. Freeman Company},\ \bibinfo {year}
  {1969})\BibitemShut {NoStop}%
\bibitem [{\citenamefont {Fowler}\ \emph {et~al.}(1981)\citenamefont {Fowler},
  \citenamefont {Raha},\ and\ \citenamefont {Weiner}}]{Fowler_Raha_Weiner}%
  \BibitemOpen
  \bibfield  {author} {\bibinfo {author} {\bibfnamefont {G.~N.}\ \bibnamefont
  {Fowler}}, \bibinfo {author} {\bibfnamefont {S.}~\bibnamefont {Raha}}, \ and\
  \bibinfo {author} {\bibfnamefont {R.~M.}\ \bibnamefont {Weiner}},\ }\href
  {\doibase 10.1007/BF01410668} {\bibfield  {journal} {\bibinfo  {journal}
  {Zeitschrift f{\"u}r Physik C Particles and Fields}\ }\textbf {\bibinfo
  {volume} {9}},\ \bibinfo {pages} {271} (\bibinfo {year} {1981})}\BibitemShut
  {NoStop}%
\bibitem [{\citenamefont {Plumer}(1984)}]{Fowler_Raha_Weiner_1}%
  \BibitemOpen
  \bibfield  {author} {\bibinfo {author} {\bibfnamefont {M.}~\bibnamefont
  {Plumer}},\ }\emph {\bibinfo {title} {Quark - Gluon - Plasma und
  vielfacherzeugung in der Starken Wechselwirkung}},\ \href@noop {} {Ph.D.
  thesis},\ \bibinfo  {school} {Philipps University at, Marburg/Lahn, Germany}
  (\bibinfo {year} {1984})\BibitemShut {NoStop}%
\bibitem [{\citenamefont {Synge}\ and\ \citenamefont {Schild}(2012)}]{synge}%
  \BibitemOpen
  \bibfield  {author} {\bibinfo {author} {\bibfnamefont {J.}~\bibnamefont
  {Synge}}\ and\ \bibinfo {author} {\bibfnamefont {A.}~\bibnamefont {Schild}},\
  }\href {https://books.google.co.in/books?id=7ey7AQAAQBAJ} {\emph {\bibinfo
  {title} {Tensor Calculus}}},\ Dover Books on Mathematics\ (\bibinfo
  {publisher} {Dover Publications},\ \bibinfo {year} {2012})\BibitemShut
  {NoStop}%
\bibitem [{\citenamefont {Hobson}\ \emph {et~al.}(2006)\citenamefont {Hobson},
  \citenamefont {P}, \citenamefont {Efstathiou},\ and\ \citenamefont
  {Lasenby}}]{hobson}%
  \BibitemOpen
  \bibfield  {author} {\bibinfo {author} {\bibfnamefont {M.}~\bibnamefont
  {Hobson}}, \bibinfo {author} {\bibfnamefont {E.}~\bibnamefont {P}}, \bibinfo
  {author} {\bibfnamefont {G.}~\bibnamefont {Efstathiou}}, \ and\ \bibinfo
  {author} {\bibfnamefont {A.}~\bibnamefont {Lasenby}},\ }\href
  {https://books.google.co.in/books?id=5dryXCWR7EIC} {\emph {\bibinfo {title}
  {General Relativity: An Introduction for Physicists}}}\ (\bibinfo
  {publisher} {Cambridge University Press},\ \bibinfo {year}
  {2006})\BibitemShut {NoStop}%
\bibitem [{\citenamefont {Landau}\ and\ \citenamefont
  {Lifshitz}()}]{gr_freq_2}%
  \BibitemOpen
  \bibfield  {author} {\bibinfo {author} {\bibfnamefont {L.~D.}\ \bibnamefont
  {Landau}}\ and\ \bibinfo {author} {\bibfnamefont {E.~M.}\ \bibnamefont
  {Lifshitz}},\ }\href@noop {} {\emph {\bibinfo {title} {The Classical The-ory
  of Fields}}}\ (\bibinfo  {publisher} {Butterworth - Heinemann, Indian Reprint
  2008})\BibitemShut {NoStop}%
\bibitem [{\citenamefont {Dubey}\ and\ \citenamefont {Sen}(2015)}]{gr_freq_1}%
  \BibitemOpen
  \bibfield  {author} {\bibinfo {author} {\bibfnamefont {A.~K.}\ \bibnamefont
  {Dubey}}\ and\ \bibinfo {author} {\bibfnamefont {A.~K.}\ \bibnamefont
  {Sen}},\ }\href {\doibase 10.1007/s10773-014-2464-3} {\bibfield  {journal}
  {\bibinfo  {journal} {Int. J. Theor. Phys.}\ }\textbf {\bibinfo {volume}
  {54}},\ \bibinfo {pages} {2398} (\bibinfo {year} {2015})}\BibitemShut
  {NoStop}%
\bibitem [{\citenamefont {Fonseca}\ \emph {et~al.}(2016)\citenamefont {Fonseca}
  \emph {et~al.}}]{J1614-2230}%
  \BibitemOpen
  \bibfield  {author} {\bibinfo {author} {\bibfnamefont {E.}~\bibnamefont
  {Fonseca}} \emph {et~al.},\ }\href {\doibase 10.3847/0004-637x/832/2/167}
  {\bibfield  {journal} {\bibinfo  {journal} {The Astrophysical Journal}\
  }\textbf {\bibinfo {volume} {832}},\ \bibinfo {pages} {167} (\bibinfo {year}
  {2016})}\BibitemShut {NoStop}%
\bibitem [{\citenamefont {Demorest}\ \emph {et~al.}(2010)\citenamefont
  {Demorest}, \citenamefont {Pennucci}, \citenamefont {Ransom}, \citenamefont
  {Roberts},\ and\ \citenamefont {Hessels}}]{Demorest}%
  \BibitemOpen
  \bibfield  {author} {\bibinfo {author} {\bibfnamefont {P.~B.}\ \bibnamefont
  {Demorest}}, \bibinfo {author} {\bibfnamefont {T.}~\bibnamefont {Pennucci}},
  \bibinfo {author} {\bibfnamefont {S.~M.}\ \bibnamefont {Ransom}}, \bibinfo
  {author} {\bibfnamefont {M.~S.~E.}\ \bibnamefont {Roberts}}, \ and\ \bibinfo
  {author} {\bibfnamefont {J.~W.~T.}\ \bibnamefont {Hessels}},\ }\href
  {\doibase 10.1038/nature09466} {\bibfield  {journal} {\bibinfo  {journal}
  {Nature}\ }\textbf {\bibinfo {volume} {467}},\ \bibinfo {pages} {1081}
  (\bibinfo {year} {2010})}\BibitemShut {NoStop}%
\bibitem [{\citenamefont {Antoniadis}\ \emph {et~al.}(2013)\citenamefont
  {Antoniadis} \emph {et~al.}}]{J0348+0432}%
  \BibitemOpen
  \bibfield  {author} {\bibinfo {author} {\bibfnamefont {J.}~\bibnamefont
  {Antoniadis}} \emph {et~al.},\ }\href {\doibase 10.1126/science.1233232}
  {\bibfield  {journal} {\bibinfo  {journal} {Science}\ }\textbf {\bibinfo
  {volume} {340}} (\bibinfo {year} {2013}),\
  10.1126/science.1233232}\BibitemShut {NoStop}%
\bibitem [{\citenamefont {Cromartie}\ \emph {et~al.}(2020)\citenamefont
  {Cromartie} \emph {et~al.}}]{J0740+6620}%
  \BibitemOpen
  \bibfield  {author} {\bibinfo {author} {\bibfnamefont {H.~T.}\ \bibnamefont
  {Cromartie}} \emph {et~al.},\ }\href {\doibase 10.1038/s41550-019-0880-2}
  {\bibfield  {journal} {\bibinfo  {journal} {Nature Astronomy}\ }\textbf
  {\bibinfo {volume} {4}},\ \bibinfo {pages} {72} (\bibinfo {year}
  {2020})}\BibitemShut {NoStop}%
\bibitem [{\citenamefont {An}\ \emph {et~al.}(2017)\citenamefont {An},
  \citenamefont {Romani}, \citenamefont {Johnson}, \citenamefont {Kerr},\ and\
  \citenamefont {Clark}}]{J1311-3430}%
  \BibitemOpen
  \bibfield  {author} {\bibinfo {author} {\bibfnamefont {H.}~\bibnamefont
  {An}}, \bibinfo {author} {\bibfnamefont {R.~W.}\ \bibnamefont {Romani}},
  \bibinfo {author} {\bibfnamefont {T.}~\bibnamefont {Johnson}}, \bibinfo
  {author} {\bibfnamefont {M.}~\bibnamefont {Kerr}}, \ and\ \bibinfo {author}
  {\bibfnamefont {C.~J.}\ \bibnamefont {Clark}},\ }\href {\doibase
  10.3847/1538-4357/aa947f} {\bibfield  {journal} {\bibinfo  {journal} {The
  Astrophysical Journal}\ }\textbf {\bibinfo {volume} {850}},\ \bibinfo {pages}
  {100} (\bibinfo {year} {2017})}\BibitemShut {NoStop}%
\bibitem [{\citenamefont {van Kerkwijk}\ \emph {et~al.}(2011)\citenamefont {van
  Kerkwijk}, \citenamefont {Breton},\ and\ \citenamefont
  {Kulkarni}}]{B1957+20}%
  \BibitemOpen
  \bibfield  {author} {\bibinfo {author} {\bibfnamefont {M.~H.}\ \bibnamefont
  {van Kerkwijk}}, \bibinfo {author} {\bibfnamefont {R.~P.}\ \bibnamefont
  {Breton}}, \ and\ \bibinfo {author} {\bibfnamefont {S.~R.}\ \bibnamefont
  {Kulkarni}},\ }\href {\doibase 10.1088/0004-637x/728/2/95} {\bibfield
  {journal} {\bibinfo  {journal} {The Astrophysical Journal}\ }\textbf
  {\bibinfo {volume} {728}},\ \bibinfo {pages} {95} (\bibinfo {year}
  {2011})}\BibitemShut {NoStop}%
\bibitem [{\citenamefont {Ord}\ \emph {et~al.}(2006)\citenamefont {Ord},
  \citenamefont {Jacoby}, \citenamefont {Hotan},\ and\ \citenamefont
  {Bailes}}]{J1600-3053_1}%
  \BibitemOpen
  \bibfield  {author} {\bibinfo {author} {\bibfnamefont {S.~M.}\ \bibnamefont
  {Ord}}, \bibinfo {author} {\bibfnamefont {B.~A.}\ \bibnamefont {Jacoby}},
  \bibinfo {author} {\bibfnamefont {A.~W.}\ \bibnamefont {Hotan}}, \ and\
  \bibinfo {author} {\bibfnamefont {M.}~\bibnamefont {Bailes}},\ }\href
  {\doibase 10.1111/j.1365-2966.2006.10646.x} {\bibfield  {journal} {\bibinfo
  {journal} {Monthly Notices of the Royal Astronomical Society}\ }\textbf
  {\bibinfo {volume} {371}},\ \bibinfo {pages} {337} (\bibinfo {year}
  {2006})}\BibitemShut {NoStop}%
\bibitem [{\citenamefont {Arzoumanian}\ \emph {et~al.}(2018)\citenamefont
  {Arzoumanian} \emph {et~al.}}]{J1600-3053}%
  \BibitemOpen
  \bibfield  {author} {\bibinfo {author} {\bibfnamefont {Z.}~\bibnamefont
  {Arzoumanian}} \emph {et~al.},\ }\href {\doibase 10.3847/1538-4365/aab5b0}
  {\bibfield  {journal} {\bibinfo  {journal} {The Astrophysical Journal
  Supplement Series}\ }\textbf {\bibinfo {volume} {235}},\ \bibinfo {pages}
  {37} (\bibinfo {year} {2018})}\BibitemShut {NoStop}%
\bibitem [{\citenamefont {Broderick}\ \emph {et~al.}(2016)\citenamefont
  {Broderick} \emph {et~al.}}]{J2215+5135_1}%
  \BibitemOpen
  \bibfield  {author} {\bibinfo {author} {\bibfnamefont {J.~W.}\ \bibnamefont
  {Broderick}} \emph {et~al.},\ }\href {\doibase 10.1093/mnras/stw794}
  {\bibfield  {journal} {\bibinfo  {journal} {Monthly Notices of the Royal
  Astronomical Society}\ }\textbf {\bibinfo {volume} {459}},\ \bibinfo {pages}
  {2681} (\bibinfo {year} {2016})}\BibitemShut {NoStop}%
\bibitem [{\citenamefont {Linares}\ \emph {et~al.}(2018)\citenamefont
  {Linares}, \citenamefont {Shahbaz},\ and\ \citenamefont
  {Casares}}]{J2215+5135}%
  \BibitemOpen
  \bibfield  {author} {\bibinfo {author} {\bibfnamefont {M.}~\bibnamefont
  {Linares}}, \bibinfo {author} {\bibfnamefont {T.}~\bibnamefont {Shahbaz}}, \
  and\ \bibinfo {author} {\bibfnamefont {J.}~\bibnamefont {Casares}},\ }\href
  {\doibase 10.3847/1538-4357/aabde6} {\bibfield  {journal} {\bibinfo
  {journal} {The Astrophysical Journal}\ }\textbf {\bibinfo {volume} {859}},\
  \bibinfo {pages} {54} (\bibinfo {year} {2018})}\BibitemShut {NoStop}%
\bibitem [{\citenamefont {Nice}\ \emph {et~al.}(2005)\citenamefont {Nice},
  \citenamefont {Splaver}, \citenamefont {Stairs}, \citenamefont {Lohmer},
  \citenamefont {Jessner}, \citenamefont {Kramer},\ and\ \citenamefont
  {Cordes}}]{J0751+1807}%
  \BibitemOpen
  \bibfield  {author} {\bibinfo {author} {\bibfnamefont {D.~J.}\ \bibnamefont
  {Nice}}, \bibinfo {author} {\bibfnamefont {E.~M.}\ \bibnamefont {Splaver}},
  \bibinfo {author} {\bibfnamefont {I.~H.}\ \bibnamefont {Stairs}}, \bibinfo
  {author} {\bibfnamefont {O.}~\bibnamefont {Lohmer}}, \bibinfo {author}
  {\bibfnamefont {A.}~\bibnamefont {Jessner}}, \bibinfo {author} {\bibfnamefont
  {M.}~\bibnamefont {Kramer}}, \ and\ \bibinfo {author} {\bibfnamefont {J.~M.}\
  \bibnamefont {Cordes}},\ }\href {\doibase 10.1086/497109} {\bibfield
  {journal} {\bibinfo  {journal} {The Astrophysical Journal}\ }\textbf
  {\bibinfo {volume} {634}},\ \bibinfo {pages} {1242} (\bibinfo {year}
  {2005})}\BibitemShut {NoStop}%
\bibitem [{\citenamefont {{Lundgren}}\ \emph {et~al.}(1995)\citenamefont
  {{Lundgren}}, \citenamefont {{Zepka}},\ and\ \citenamefont
  {{Cordes}}}]{J0751+1807_1}%
  \BibitemOpen
  \bibfield  {author} {\bibinfo {author} {\bibfnamefont {S.~C.}\ \bibnamefont
  {{Lundgren}}}, \bibinfo {author} {\bibfnamefont {A.~F.}\ \bibnamefont
  {{Zepka}}}, \ and\ \bibinfo {author} {\bibfnamefont {J.~M.}\ \bibnamefont
  {{Cordes}}},\ }\href {\doibase 10.1086/176402} {\bibfield  {journal}
  {\bibinfo  {journal} {\apj}\ }\textbf {\bibinfo {volume} {453}},\ \bibinfo
  {pages} {419} (\bibinfo {year} {1995})}\BibitemShut {NoStop}%
\bibitem [{\citenamefont {Freire}\ \emph {et~al.}(2008)\citenamefont {Freire},
  \citenamefont {Bassa}, \citenamefont {Wang}, \citenamefont {Cumming},\ and\
  \citenamefont {Kaspi}}]{B1516+02B}%
  \BibitemOpen
  \bibfield  {author} {\bibinfo {author} {\bibfnamefont {P.~C.~C.}\
  \bibnamefont {Freire}}, \bibinfo {author} {\bibfnamefont {C.}~\bibnamefont
  {Bassa}}, \bibinfo {author} {\bibfnamefont {Z.}~\bibnamefont {Wang}},
  \bibinfo {author} {\bibfnamefont {A.}~\bibnamefont {Cumming}}, \ and\
  \bibinfo {author} {\bibfnamefont {V.~M.}\ \bibnamefont {Kaspi}},\ }\href
  {\doibase 10.1063/1.2900274} {\bibfield  {journal} {\bibinfo  {journal} {AIP
  Conference Proceedings}\ } (\bibinfo {year} {2008}),\
  10.1063/1.2900274}\BibitemShut {NoStop}%
\bibitem [{\citenamefont {Weber}(2005)}]{bag145}%
  \BibitemOpen
  \bibfield  {author} {\bibinfo {author} {\bibfnamefont {F.}~\bibnamefont
  {Weber}},\ }\href {\doibase 10.1016/j.ppnp.2004.07.001} {\bibfield  {journal}
  {\bibinfo  {journal} {Prog. Part. Nucl. Phys.}\ }\textbf {\bibinfo {volume}
  {54}},\ \bibinfo {pages} {193} (\bibinfo {year} {2005})}\BibitemShut
  {NoStop}%
\bibitem [{\citenamefont {Banerjee}(2018)}]{lim_bag}%
  \BibitemOpen
  \bibfield  {author} {\bibinfo {author} {\bibfnamefont {S.}~\bibnamefont
  {Banerjee}},\ }\href {\doibase 10.1007/s12043-018-1597-y} {\bibfield
  {journal} {\bibinfo  {journal} {Pramana}\ }\textbf {\bibinfo {volume} {91}},\
  \bibinfo {pages} {27} (\bibinfo {year} {2018})}\BibitemShut {NoStop}%
\bibitem [{\citenamefont {{Manchester}}\ \emph {et~al.}(2005)\citenamefont
  {{Manchester}}, \citenamefont {{Hobbs}}, \citenamefont {{Teoh}},\ and\
  \citenamefont {{Hobbs}}}]{atnf}%
  \BibitemOpen
  \bibfield  {author} {\bibinfo {author} {\bibfnamefont {R.~N.}\ \bibnamefont
  {{Manchester}}}, \bibinfo {author} {\bibfnamefont {G.~B.}\ \bibnamefont
  {{Hobbs}}}, \bibinfo {author} {\bibfnamefont {A.}~\bibnamefont {{Teoh}}}, \
  and\ \bibinfo {author} {\bibfnamefont {M.}~\bibnamefont {{Hobbs}}},\ }\href
  {\doibase 10.1086/428488} {\bibfield  {journal} {\bibinfo  {journal} {The
  Astronomical Journal}\ }\textbf {\bibinfo {volume} {129}},\ \bibinfo {pages}
  {1993} (\bibinfo {year} {2005})}\BibitemShut {NoStop}%
\bibitem [{\citenamefont {Gourgoulhon}\ \emph {et~al.}(1999)\citenamefont
  {Gourgoulhon}, \citenamefont {Haensel}, \citenamefont {Livine}, \citenamefont
  {Paluch}, \citenamefont {Bonazzola},\ and\ \citenamefont
  {Marck}}]{upperlim_fr}%
  \BibitemOpen
  \bibfield  {author} {\bibinfo {author} {\bibfnamefont {E.}~\bibnamefont
  {Gourgoulhon}}, \bibinfo {author} {\bibfnamefont {P.}~\bibnamefont
  {Haensel}}, \bibinfo {author} {\bibfnamefont {R.}~\bibnamefont {Livine}},
  \bibinfo {author} {\bibfnamefont {E.}~\bibnamefont {Paluch}}, \bibinfo
  {author} {\bibfnamefont {S.}~\bibnamefont {Bonazzola}}, \ and\ \bibinfo
  {author} {\bibfnamefont {J.}~\bibnamefont {Marck}},\ }\href@noop {}
  {\bibfield  {journal} {\bibinfo  {journal} {Astron. Astrophys.}\ }\textbf
  {\bibinfo {volume} {349}},\ \bibinfo {pages} {851} (\bibinfo {year}
  {1999})},\ \Eprint {http://arxiv.org/abs/astro-ph/9907225}
  {arXiv:astro-ph/9907225} \BibitemShut {NoStop}%
\bibitem [{\citenamefont {{Singh}}\ \emph {et~al.}(2020)\citenamefont
  {{Singh}}, \citenamefont {{Mallick}},\ and\ \citenamefont
  {{Prasad}}}]{ssrmrp}%
  \BibitemOpen
  \bibfield  {author} {\bibinfo {author} {\bibfnamefont {S.}~\bibnamefont
  {{Singh}}}, \bibinfo {author} {\bibfnamefont {R.}~\bibnamefont {{Mallick}}},
  \ and\ \bibinfo {author} {\bibfnamefont {R.}~\bibnamefont {{Prasad}}},\
  }\href@noop {} {\bibfield  {journal} {\bibinfo  {journal} {arXiv e-prints}\
  ,\ \bibinfo {eid} {arXiv:2003.00693}} (\bibinfo {year} {2020})},\ \Eprint
  {http://arxiv.org/abs/2003.00693} {arXiv:2003.00693 [astro-ph.HE]}
  \BibitemShut {NoStop}%
\end{thebibliography}%

\end{document}